\documentclass[prb,amsmath,amssymb,superscriptaddress, twocolumn]{revtex4}
\usepackage{float}
\usepackage{amsmath}
\usepackage{amssymb}
\usepackage{amsfonts}
\usepackage{euscript}
\usepackage{enumerate}
\usepackage{hhline}
\usepackage{pslatex}
\usepackage{tabularx}
\usepackage[usenames,dvipsnames]{xcolor}

\usepackage{graphicx}

\usepackage{dcolumn}
\usepackage{bm}
\usepackage[sort&compress]{natbib}

\makeatletter
\renewcommand{\@biblabel}[1]{#1. }
\renewcommand{\@dotsep}{500}
\renewcommand{\@pnumwidth}{0em}
\renewcommand{\l@figure}[2]{
\@dottedtocline{1}{1.5em}{2em}{Figure #1}{}\vspace{15pt}}

\usepackage[normalem]{ulem}

\begin{document}

\title{Efficient photo-induced second harmonic generation in silicon photonics}

\author{Xiyuan Lu}\email{xiyuan.lu@nist.gov}
\affiliation{Microsystems and Nanotechnology Division, Physical Measurement Laboratory, National Institute of Standards and Technology, Gaithersburg, MD 20899, USA}
\affiliation{Institute for Research in Electronics and Applied Physics and Maryland NanoCenter, University of Maryland,
College Park, MD 20742, USA}
\author{Gregory Moille}
\affiliation{Microsystems and Nanotechnology Division, Physical Measurement Laboratory, National Institute of Standards and Technology, Gaithersburg, MD 20899, USA}
\affiliation{Joint Quantum Institute, NIST/University of Maryland,
College Park, MD 20742, USA}
\author{Ashutosh Rao}
\affiliation{Microsystems and Nanotechnology Division, Physical Measurement Laboratory, National Institute of Standards and Technology, Gaithersburg, MD 20899, USA}
\affiliation{Department of Chemistry and Biochemistry, University of Maryland,
College Park, MD 20742, USA}
\author{Daron A. Westly}
\affiliation{Microsystems and Nanotechnology Division, Physical Measurement Laboratory, National Institute of Standards and Technology, Gaithersburg, MD 20899, USA}
\author{Kartik Srinivasan} \email{kartik.srinivasan@nist.gov}
\affiliation{Microsystems and Nanotechnology Division, Physical Measurement Laboratory, National Institute of Standards and Technology, Gaithersburg, MD 20899, USA}
\affiliation{Joint Quantum Institute, NIST/University of Maryland, College Park, MD 20742, USA}

\date{\today}

\begin{abstract}
        \noindent Silicon photonics lacks a second-order nonlinear optical ($\chi^{(2)}$) response in general because the typical constituent materials are centro-symmetric  and lack inversion symmetry, which prohibits $\chi^{(2)}$ nonlinear processes such as second harmonic generation (SHG). Here, we realize record-high SHG efficiency in silicon photonics by combining a photo-induced effective $\chi^{(2)}$ nonlinearity with resonant enhancement and perfect-phase matching. We show a conversion efficiency of (2,500 $\pm$ 100) \%/W, which is 2 to 4 orders of magnitude larger than previous works. In particular, our devices realize mW-level SHG output powers with $>$ 20~$\%$ power conversion efficiency. This demonstration is a major breakthrough in realizing efficient $\chi^{(2)}$ processes in silicon photonics, and paves the way for integrated self-referencing of Kerr frequency combs for compact optical frequency synthesis and optical clock technologies.
\end{abstract}

\maketitle
\noindent Second-order ($\chi^{(2)}$) nonlinear optical processes are a cornerstone for many classical and quantum applications~\cite{Boyd2008}. For example, to achieve compact functionalities for optical frequency synthesis~\cite{Spencer_Nature_2018, Singh_CLEO_2019} and optical atomic clocks~\cite{Newman_Optica_2019}, frequency combs based on third-order nonlinear processes need to be self-referenced by efficient second harmonic generation (SHG), ideally on the same silicon chip. However, common materials in silicon photonics, including silicon (Si), silicon nitride (Si$_3$N$_4$), and silicon dioxide (SiO$_2$), do not support $\chi^{(2)}$ response in bulk within the electric-dipole approximation~\cite{Boyd2008}. It is therefore particularly challenging to realize efficient SHG on a silicon chip. As a result, systems based on silicon photonics technology have often relied on conventional platforms such as centimeter-scale periodically-poled lithium niobate waveguides for $\chi^{(2)}$  functionalities~\cite{Spencer_Nature_2018,Newman_Optica_2019}. Alternatively, there has been considerable progress in realizing efficient SHG in non-silicon-based thin film platforms, including aluminum nitride \cite{Guo_Optica_2016}, gallium arsenide \cite{Lin_APLPhoton_2019}, and lithium niobate \cite{Juanjuan_Optica_2019,Wang_Optica_2018, Luo_Optica_2018, Lin_Optica_2016}. Such advances generally require heterogeneous integration with a silicon-based platform~\cite{Lin_OL_2017} for optimized performance in the aforementioned frequency comb applications. On the other hand, silicon carbide nanophotonics has recently made major strides, demonstrating $\chi^{(2)}$ processes in both photonic crystal cavities~\cite{Noda_Optica_2019} and microring resonators~\cite{Lukin_NatPhoton_2019}. However, the fabrication processes that realize high performance in SiC~\cite{Noda_Optica_2019,Lukin_NatPhoton_2019} suggest that its integration with the rest of the silicon photonics platform may be challenging.

There has also been work aiming for demonstrating effective $\chi^{(2)}$ processes directly in typical silicon photonics materials. One approach uses the weak $\chi^{(2)}$ nonlinearity present in silicon-based systems (for example, due to symmetry breaking) in conjunction with perfect-phase matching in high quality factor ($Q$) microcavities to boost the normalized SHG efficiencies to 0.1~\%/W~[\onlinecite{Levy_OE_2011}] and 0.049~\%/W~[\onlinecite{Zhang_NatPhoton_2019}]. These efficiencies can be improved by optimized input/output waveguide-resonator coupling, but are ultimately limited by the weakness of the $\chi^{(2)}$ nonlinearity. Another approach uses a large effective $\chi^{(2)}$ nonlinearity created through the combination of an electric field and the medium's $\chi^{(3)}$ nonlinearity in photonic waveguides without cavity enhancement. This electric field can be induced directly by external electrodes~\cite{Timurdogan_NatPhoton_2017} or optically through the photo-galvanic effect\cite{Porcel_OE_2017, Billat_NatCommun_2017, Hickstein_NatPhoton_2019, Grassani_OL_2019,Edgars_ACSPhoton_2019}, and has yielded $\chi^{(2)}$ nonlinearities ranging from 0.3 pm/V to 3.7 pm/V, resulting in normalized SHG efficiencies as high as 13~\%/W~~[\onlinecite{Timurdogan_NatPhoton_2017}]. The induced field not only produces a nonlinearity that significantly exceeds the existing intrinsic nonlinearty, but also supports quasi-phase-matching, with phase either pre-determined~\cite{Timurdogan_NatPhoton_2017} or self-organized/photo-induced~\cite{Billat_NatCommun_2017, Hickstein_NatPhoton_2019, Edgars_ACSPhoton_2019}. However, the above approaches, when used separately, are inefficient compared to devices using traditional $\chi^{(2)}$ materials~\cite{Guo_Optica_2016, Lin_APLPhoton_2019, Juanjuan_Optica_2019}, and as a result are far from generating~mW-level continuous-wave SHG output.

\begin{figure*}[t!]
\centering\includegraphics[width=0.88\linewidth]{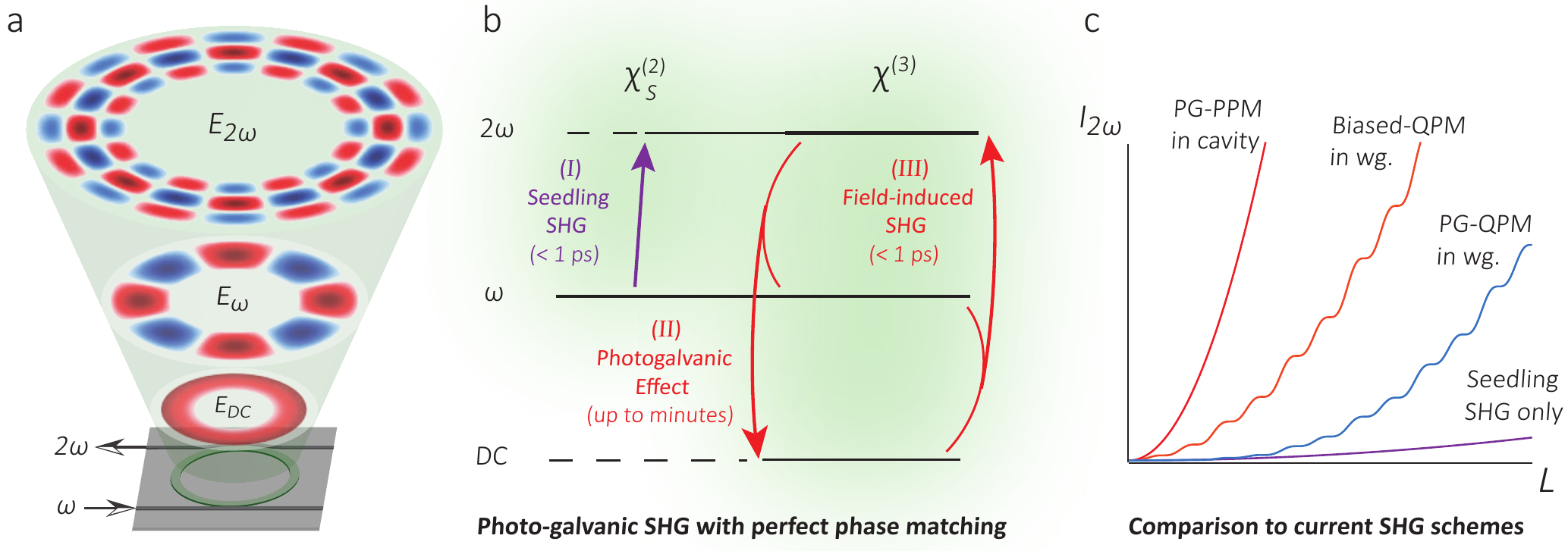}
\caption{\textbf{Photo-induced second harmonic generation (SHG) in a Si$_3$N$_4$ microring resonator.} \textbf{a,} Illustration of the device and modes involved in perfect-phase matching. A Si$_3$N$_4$ microring is integrated with two coupling waveguides for pump ($\omega$) and SHG ($2\omega$) light. Within the microring, the mode profiles of three interacting modes at $2\omega$, $\omega$, and DC frequencies are shown from top to bottom. Red and blue indicate positive and negative phase of the electric fields, that is, pointing outwards or inwards in radial direction. The darkness of the colors depicts the local field strength. These modes satisfy perfect-phase matching for both $\chi^{(2)}$ and $\chi^{(3)}$ processes. \textbf{b,} Photogalvanic SHG with perfect-phase matching in a microring. (I) A small seedling SHG ($2\omega$) field is generated with a weak intrinsic $\chi^{(2)}$ nonlinearity. (II) This $2\omega$ field, together with the $\omega$ field, builds up a DC field through the coherent photogalvanic effect, which is distinctively slow (up to minutes) compared to ordinary nonlinear optical processes ($<$ 1 ps). (III) The generated DC field and the pump field in turn generate SHG light through a field-induced SHG effect, i.e., DC Kerr effect. Once seeded by (I), (II) and (III) work together to build up the SHG field. \textbf{c,} Our SHG scheme (red) is superior in power efficiency than other processes in $\chi^{(2)}$ silicon photonics, including biased-QPM \cite{Timurdogan_NatPhoton_2017} (orange), photogalvanic-QPM \cite{Billat_NatCommun_2017, Hickstein_NatPhoton_2019, Edgars_ACSPhoton_2019} (blue), and high-Q resonators with weak nonlinearities \cite{Levy_OE_2011,Zhang_NatPhoton_2019} (purple). See the Supplementary for more details. PG: photogalvanic. PPM: perfect-phase matching. QPM: quasi-phase matching. wg.: waveguide.}
\label{Fig1}
\vspace{-12 pt}
\end{figure*}

To realize efficient SHG, we engineer devices that take advantage of both a strong effective $\chi^{(2)}$ nonlinearity and resonant enhancement. We use the Si$_3$N$_4$ platform~\cite{Moss_NatPhoton_2013} that has been successfully applied to many wide-band nonlinear photonics demonstrations, including octave-spanning frequency combs~\cite{Okawachi_OL_2011, Li_Optica_2017, Karpov_NatCommun_2018}, classical/quantum frequency conversion~\cite{Li_NatPhoton_2016, Lu_NatPhoton_2019}, and optical parametric oscillation~\cite{Lu_Optica_2019}. The physical process is photogalvanic field-induced SHG, first discovered in germanium-doped glass fibers decades ago~\cite{UO_OL_1986}. In contrast to reports of a photo-induced $\chi^{(2)}$ in non-resonant geometries such as Si$_3$N$_4$ waveguides~\cite{Porcel_OE_2017, Billat_NatCommun_2017, Hickstein_NatPhoton_2019, Grassani_OL_2019, Edgars_ACSPhoton_2019} and SiO$_2$ fibers~\cite{UO_OL_1986, UO_OL_1987, Tom_OL_1987,Tom_OL_1988,Margulis_Nature_1995}, here we demonstrate an effective photo-induced $\chi^{(2)}$ nonlinearity in a high-$Q$ Si$_3$N$_4$ microresonator. We show that this resonantly-enhanced, field-induced $\chi^{(2)}$ nonlinear process enables high efficiency SHG with appreciable output power for continuous wave inputs.

The physical process behind our approach involves $\chi^{(2)}$ and $\chi^{(3)}$ nonlinear interaction among three modes, as shown in Fig.~\ref{Fig1}(a,b), and is unique in realizing perfect-phase matching for both the intrinsic $\chi^{(2)}$ process and the field-induced $\chi^{(2)}$ process simultaneously. First, phase and frequency matching of the intrinsic $\chi^{(2)}$ process, as shown in Fig.~\ref{Fig1}(b)-(I), is supported by recent techniques developed in nanophotonic dispersion engineering~\cite{Lu_NatPhoton_2019}. Once such matching is fulfilled, the $\chi^{(3)}$ processes, as shown in Fig.~\ref{Fig1}(b)-(II,III), are automatically matched given the nature of the DC field, which is stationary with zero angular momentum and has a frequency of $\omega = 0$. The $\chi^{(2)}$ and $\chi^{(3)}$ processes can therefore work together seamlessly, that is, the induced $\chi^{(2)}$ can feed upon the intrinsic seedling SHG to self-start (Fig.~\ref{Fig1}(b)), rather than relying on external electrodes \cite{Timurdogan_NatPhoton_2017} or a SHG laser for initiation~\cite{Billat_NatCommun_2017}. Second, the field-induced $\chi^{(2)}$ in our scheme, through perfect-phase matching with a DC field, is more efficient than those reported previously, which are achieved through quasi~phase~matching with RF fields \cite{Timurdogan_NatPhoton_2017, Billat_NatCommun_2017, Hickstein_NatPhoton_2019, Edgars_ACSPhoton_2019}, as shown in Fig.~\ref{Fig1}(c). Due to such perfect-phase matching, our induced $\chi^{(2)}$ from the DC field is always at its maximum, instead of having periodic modulations which decrease the effective $\chi^{(2)}$ to 2/$\pi$ and 1/$\pi$ times the maximum value for square\cite{Timurdogan_NatPhoton_2017} and sinusoidal\cite{Billat_NatCommun_2017, Hickstein_NatPhoton_2019, Edgars_ACSPhoton_2019} longitudinal profiles of the RF fields, respectively. Moreover, the resonance nature enables the induced $\chi^{(2)}$ to remain spatially uniform inside the resonator, instead of forming build-up and decay-down regions as observed in waveguide geometries~\cite{Hickstein_NatPhoton_2019,Billat_NatCommun_2017, Edgars_ACSPhoton_2019}. Through this phase-matched, photo-induced, and resonant SHG process, we report a SHG conversion efficiency of (2,500 $\pm$ 100)~\%/W in a Si$_3$N$_4$ microring resonator, which is 2 to 4 orders of magnitudes larger than previous works in typical silicon-based materials \cite{Timurdogan_NatPhoton_2017,Levy_OE_2011,Billat_NatCommun_2017,Zhang_NatPhoton_2019,Hickstein_NatPhoton_2019,Ning_OL_2012,Porcel_OE_2017, Grassani_OL_2019, Edgars_ACSPhoton_2019}. Our absolute SHG efficiency is (22~$\pm$~1)~$\%$ at an output power of (1.9~$\pm$~0.1)~mW, with an input pump power of (8.8~$\pm$~1.0)~mW. This performance level is suitable for $f$-2$f$ self-referencing of octave-spanning microresonator frequency combs within a common Si$_3$N$_4$ platform~\cite{Li_Optica_2017,Karpov_NatCommun_2018,Spencer_Nature_2018,Newman_Optica_2019}.

\begin{figure*}[t!]
\centering\includegraphics[width=0.88\linewidth]{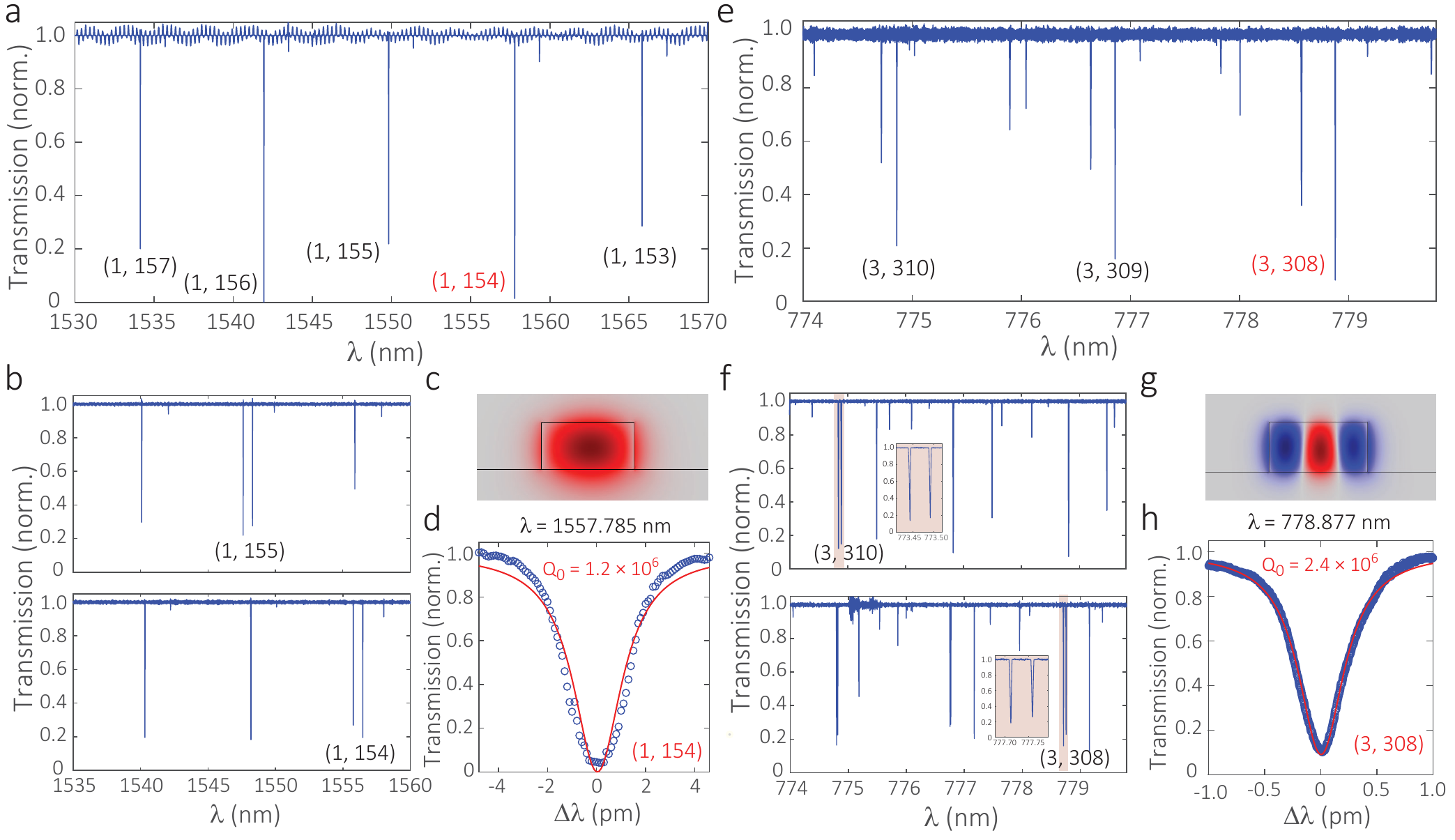}
\caption{\textbf{Device transmission shows perfect-phase matching with high-Q resonances.} Pump (\textbf{a-d}) and SHG (\textbf{e-h}) mode characteristics. \textbf{a,~e,} Normalized cavity transmission ($T$) in the 1550~nm band and the 780~nm band, respectively. The modes are labeled as ($n$,$m$), where $n$ and $m$ represent radial and azimuthal mode numbers. In particular, the (1, 154) and (3,308) modes (labeled in red) satisfy perfect-phase matching and are used in experiment as pump and SHG modes. \textbf{b,} The pump mode number is identified using mode splitting devices targeting (1, 155) and (1, 154), respectively, and the corresponding cavity modes are clearly identified by mode splitting of $\approx$ 0.7~nm. \textbf{f,} For 780~nm band modes, that is, (3, 310) and (3, 308) modes, the mode splittings are $\approx$ 45~pm, as shown in the insets. \textbf{c,~g,} Cross-sectional field profiles for the pump and SHG modes. \textbf{d,~h,} The pump and SHG modes, that is, (1, 154) and (3, 308), have wavelengths of 1557.785 nm and 778.877~nm at 21.9~$^\text{o}$C in the cold cavity (without Kerr shift or thermal bistability). Their intrinsic Q values are $(1.2 \pm 0.1) \times 10^6$  and $(2.4 \pm 0.1) \times 10^6$, and the loaded Q values are (6.0 $\pm$ 0.5) $\times 10^5$ and (1.6 $\pm$ 0.1) $\times 10^6$, respectively. The errors represent one-standard deviation uncertainties in nonlinear fitting of the resonances.}
\label{Fig2}
\vspace{-12 pt}
\end{figure*}

The SHG device is a Si$_3$N$_4$ microring integrated with two coupling waveguides, as shown in Fig.~\ref{Fig1}(a). A pump laser in the telecom band ($\approx$ 1560~nm) is coupled by a straight waveguide (bottom) into the microring, in which the 1560~nm light is frequency doubled to 780~nm, and the 780~nm light is out-coupled by a separate waveguide (top). This top waveguide only supports the 780~nm light, and does not support any modes for 1560~nm~(see Supplementary for more details on coupling). Inside the microring, three modes are involved in the process, the DC mode, the pump mode (1560~nm), and the SHG mode (780~nm). The DC mode is stationary (resonance frequency of $\omega = 0$) with zero azimuthal angular momentum, and can be characterized by ($n$, $m$) = (0, 0), when depicted in the whispering-gallery-mode terminology, where $n$ represents radial mode number and $m$ represents azimuthal mode number. To perfectly match the mode numbers for the pump and SHG, we use the fundamental transverse-electric (TE1) mode for the pump and the third-order transverse-electric (TE3) mode for the second harmonic, with the respective mode profiles in Fig.~\ref{Fig2}(c,g). Finite-element-method simulation indicates that the (1, 154) and (3, 308) modes, which are clearly matched in angular momentum, have nearly matched frequencies in a microring with 23~$\mu$m radius, 1.2~$\mu$m ring width, and 600~nm thickness~(additional details regarding dispersion engineering are in the Supplementary).

To guarantee perfect-phase matching of the underlying intrinsic $\chi^{(2)}$ process, we identify the azimuthal mode numbers using the selective mode splitting method~\cite{Lu_APL_2014} that has recently been applied to wide-band microcavity nonlinear optics~\cite{Lu_NatPhoton_2019}. With this method, we pattern some devices with a small modulation in ring width (amplitude of 20 nm) and an angular period of $\pi$/$m$ to deterministically scatter and split the $m$-th mode, while leaving other modes unperturbed, as illustrated in Fig.~\ref{Fig2}(b,f). For example, spectra from two devices exhibiting a targeted mode splitting at $\approx$ 1548~nm ($m$ = 155) and $\approx$ 1557~nm ($m$ = 154) are shown in Fig.~\ref{Fig2}(b). We also use this method to identify the 780~nm modes, as shown in Fig.~\ref{Fig2}(f). Devices on the same chip with nominal dimensions, but without the mode splitting pattern, show very similar resonance wavelengths (Fig.~\ref{Fig2}(a,~e)), so relative modes for all devices on the chip can be identified.

\begin{figure*}[t!]
\centering\includegraphics[width=0.88\linewidth]{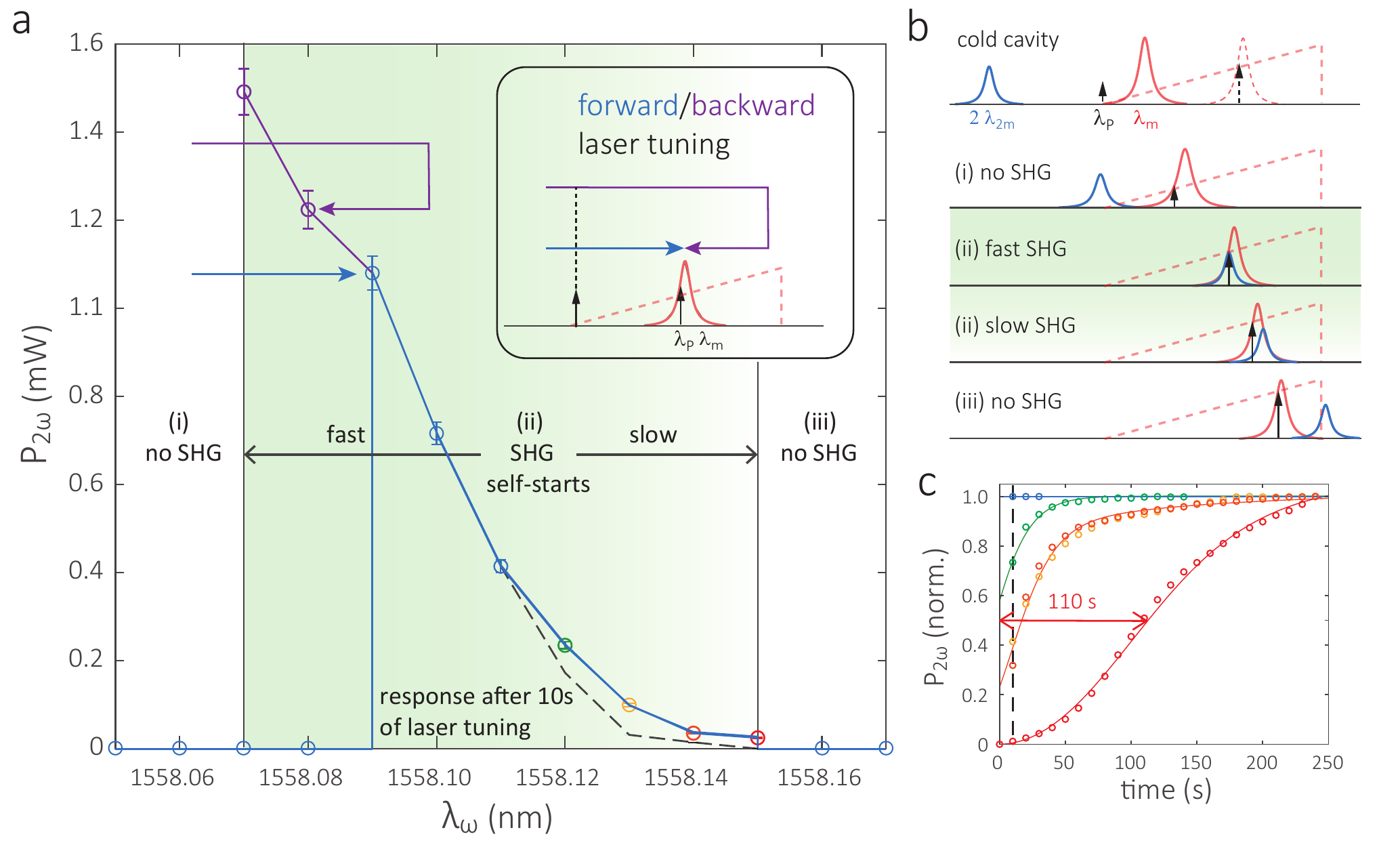}
\caption{\textbf{Laser detuning to optimize photo-induced SHG in the microring.} \textbf{a, b,} Experimental results (a) and illustrations (b) show three regions with distinct SHG response. In regions (i) and (iii), the device has no observed SHG response, because of the inefficient seedling SHG process, as the cavity modes are frequency mismatched. In region (ii) (green area), the cavity frequencies match and photo-induced SHG can self-start, either through forward (blue) or backward (red) laser tuning. As shown in the inset, both tuning methods requires forward tuning at first to drop laser power into the cavity, because of thermal bistability indicated by the dashed red line, whose height is proportional to the pump power that is dropped into the cavity. In region (ii), photo-induced SHG can self-start and the response time depends on the laser detuning, which affects the cavity frequency mismatch, as shown in (b)-(ii). Darker green indicates a faster response in region (ii). The dashed black line in (a) indicates the output SHG power after the laser is tuned for 10 seconds. The errorbars in (a) represent one-standard deviation uncertainties from the calibration of the on-chip power. \textbf{c,} The response takes a few seconds or less than a second when the pump is below 1558.11 nm, and takes from 20 seconds at 1558.12 nm (green) to 110 seconds at 1558.15 nm (red). This slow response is a signature of the photogalvanic process, in contrast to other nonlinear processes that are typically ultrafast ($<$ 1 ps). The dashed black line indicates a time of 10 seconds after laser tuning, corresponding to that in (a).}
\label{Fig3}
\vspace{-12 pt}
\end{figure*}

While phase matching is quantized and is perfect once the appropriate modes are identified, frequency matching of those modes typically needs thermal/power tuning~\cite{Guo_Optica_2016}, as shown in Fig.~\ref{Fig3}(a,b). The device has loaded quality factors ($Q$s) of $>$ 0.5 $\times10^6$ and intrinsic $Q$s $>$ 1 $\times10^6$ for both pump and SHG modes (Fig.~\ref{Fig2}(d,h)). These high $Q$s necessitate frequency matching to be within $\approx$~0.3 GHz (the cavity linewidth). The resonance wavelengths of the pump and SHG modes are 1557.785 nm and 778.877 nm, respectively, recorded by a wavemeter at room temperature, when optical power is small so that both Kerr and thermo-optic shifts are negligible. The SHG mode thus needs a $\approx$ 7.7~GHz (15.5~pm) red-shift relative to that of the pump mode to enable frequency matching, as illustrated in the top panel of Fig.~\ref{Fig3}(b). Such a frequency mismatch can be compensated by both thermal and Kerr effects. In particular, the thermal shift can make up for this frequency mismatch by heating at a rate of 0.291~GHz/$^\text{o}$C (0.585~pm/$^\text{o}$C) (see the Supplementary for details).

The thermo-optic bistability exhibited by the high-Q cavity, indicated by dashed red triangles in Fig.~\ref{Fig3}(a,b), requires the pump laser to be scanned from blue-detuning to red-detuning to drop power into the cavity. When pump power is first dropped into the cavity, illustrated by region (i) in Fig.~\ref{Fig3}(a,b), the pump and SHG mode are mismatched in frequency, similar to the cold cavity case. Here the seedling SHG process is only resonantly enhanced by the pump cavity mode but not by the SHG cavity mode, and yields no observable SHG signal (i.e., $P_\text{SHG}<$ 0.1 nW). Without such seedling SHG, effective photo-induced SHG cannot self-start. When the pump laser is tuned into region (ii) in Fig.~\ref{Fig3}(a,b), the two cavity modes start to have spectral overlap, which results in an appreciable seedling SHG power to start the photogalvanic effect.

To reach optimal SHG power, both forward and  backward tuning of the pump laser are required, with the specifics dependent on the laser-cavity detuning. These two tuning methods are illustrated in the inset of Fig.~\ref{Fig3}(a). For example, the SHG power of 1.15~mW can be directly generated by forward tuning, where the laser is tuned from $<$~1558.06 nm to 1558.09 nm, as indicated by the blue arrow. But the larger SHG power of $\approx$ 1.2~mW at 1558.08 nm can only be accessed through backward tuning, that is, the laser is first tuned in the forward direction from $<$~1558.06 nm to $>$~1558.09 nm, and then tuned in the backward direction to 1558.08 nm, as indicated by the red arrow. In comparison, when the laser is directly forward tuned from $<$~1558.06 nm to 1558.08 nm, no SHG signal is observed. Such a hysteresis is likely due to pump depletion, as the cavity frequency matching is different when SHG just starts (without depletion) in comparison to when it has already started (with depletion).

\begin{figure*}[t!]
\centering\includegraphics[width=0.88\linewidth]{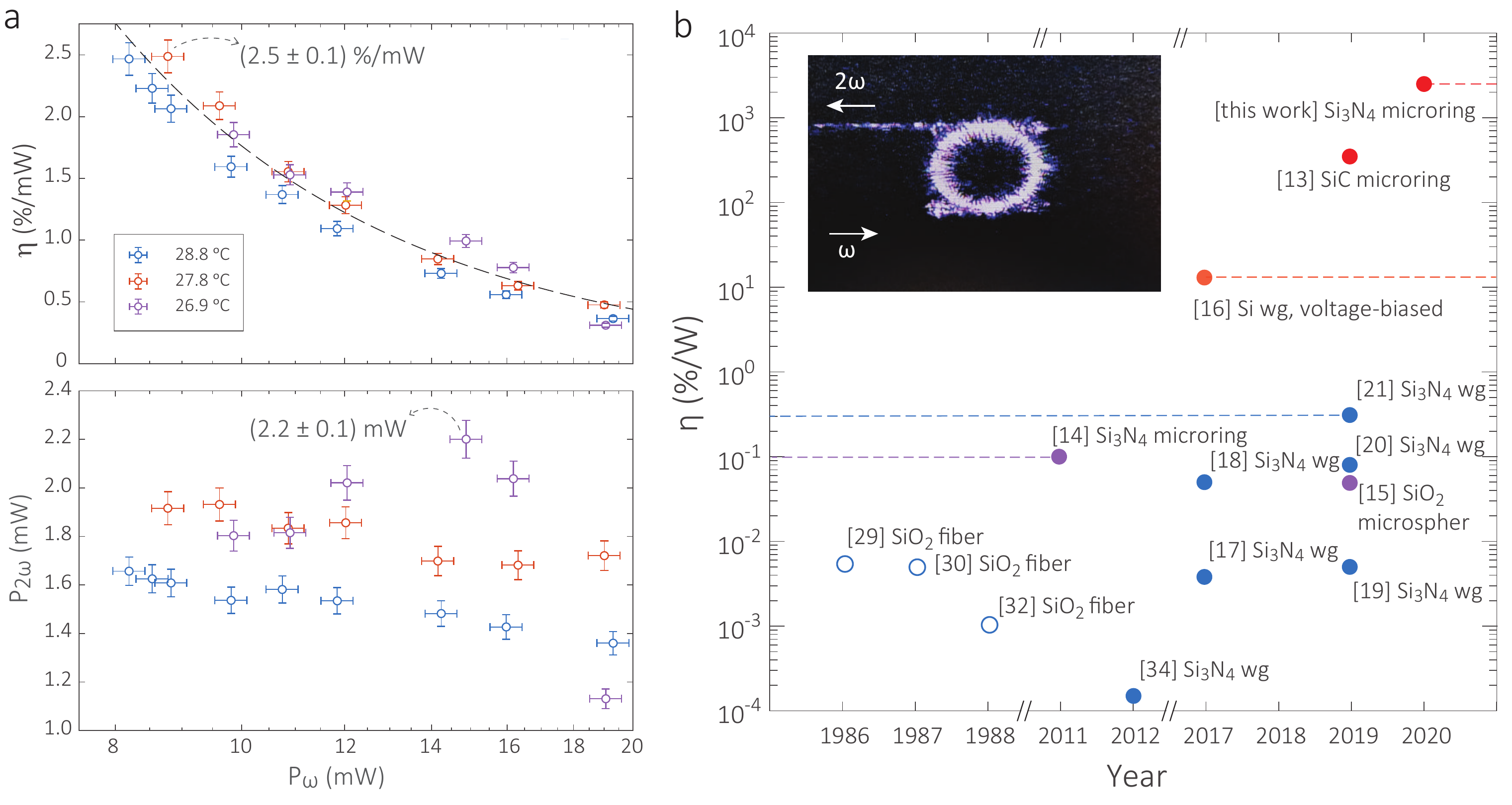}
\caption{\textbf{Record-high efficiency is achieved by photo-induced SHG in silicon photonics.} \textbf{a,} An efficiency of (2,500 $\pm$ 100)~\%/W is obtained with the laser detuning and thermal tuning (top). Here efficiency is given by $\eta = P_{\omega}/P_{2\omega}^2$, where $P_{\omega}$ and $P_{2\omega}$ represent pump and SHG powers in the waveguides on chip, respectively. The SHG power is (2.2 $\pm$ 0.1)~mW when pumped with 15~mW at 26.9 $^\text{o}$C (bottom). The error bars represent one-standard deviation uncertainties from the calibration of the on-chip power. \textbf{b,} A comparison with other SHG works in silicon photonics shows that our SHG efficiency is record-high. Inset shows the generation of SHG light by a telecom pump. The CMOS camera is responsive to 780~nm light but not telecom light. The image clearly shows the generation of SHG light in the microring and the coupling of light into the top waveguide, as indicated by the arrow. wg: waveguide.}
\label{Fig4}
\vspace{-12 pt}
\end{figure*}

As discussed earlier, when the pump laser is set to a wavelength between 1558.09 nm and 1558.15 nm, SHG can self-start simply by forward tuning. The response time of the process is determined by the photogalvanic process. This time critically depends on the cavity frequency matching. When the frequency is matched well (the top panel (ii) of Fig.~\ref{Fig3}(b)), i.e., $\lambda_\textrm{p}$ = 1558.09 nm to 1558.11 nm, the response time is within a few seconds. In contrast, when the cavity modes are not well frequency matched (the bottom panel (ii) of Fig.~\ref{Fig3}(b)), i.e., $\lambda_\textrm{p}$ = 1558.12 nm to 1558.15 nm, the SHG has a slow build-up time ranging from 20 seconds to over a minute, as shown in Fig.~\ref{Fig3}(c). Note that our photogalvanic build-up time is much shorter than those reported in previous works. Even the slowest case has a response of $\approx$ 110~s, and this response time is in general 2 to 3 orders of magnitude faster than previous photogalvanic work in waveguides \cite{Billat_NatCommun_2017, Hickstein_NatPhoton_2019, Edgars_ACSPhoton_2019}. Such a significant change in response time is related to the cavity enhancement in the SHG efficiency. While in the waveguides the SHG signal has to build up longitudinally, the SHG signal can pass through the cavity $\approx$ 5,000 times to build up coherently. This number of roundtrips of light is estimated by  $\mathcal{F}$/(2$\pi$), where $\mathcal{F}$ is the cavity finesse. Finally, when the laser is further detuned into region (iii) in Fig.~\ref{Fig3}(a,b), the cavity frequency mismatch is over-compensated, so that the SHG mode is red-shifted relative to the pump mode.

We repeat this method at various pump powers and temperatures to optimize the normalized conversion efficiency ($\eta$)  and output power ($P_{2\omega}$), as shown in Fig.~\ref{Fig4}(a). The top panel shows that the maximal normalized conversion efficiency ($\eta = P_{2\omega}/P_{\omega}$) is (2,500 $\pm$ 100)~\%/W. In comparison, as shown in Fig.~\ref{Fig4}(b), this efficiency (dashed red line) is $>100\times$ the previous record 13~\%/W in silicon photonics~\cite{Timurdogan_NatPhoton_2017} (dashed orange line). Our results have superior efficiency than previous works due to the use of perfect-phase-matching and resonant enhancement in comparison to quasi-phase-matching in waveguides~\cite{Porcel_OE_2017, Ning_OL_2012,Grassani_OL_2019, Hickstein_NatPhoton_2019,Billat_NatCommun_2017,Timurdogan_NatPhoton_2017, Edgars_ACSPhoton_2019} (dashed blue line) and silica fibers \cite{UO_OL_1986,UO_OL_1987,Tom_OL_1987,Tom_OL_1988}, while our much larger nonlinearity leads to improved performance relative to all other resonant schemes in silicon photonics \cite{Levy_OE_2011,Zhang_NatPhoton_2019} (dashed purple line).

Although our normalized SHG efficiency of (2,500 $\pm$ 100)~\%/W is only 1~\% of the recently-reported $\chi^{(2)}$ record in lithium niobate microrings~\cite{Juanjuan_Optica_2019}, our absolute SHG efficiency of (22 $\pm$ 1) \% (see Supplementary Material) is comparable to state-of-the-art $\chi^{(2)}$ nanophotonic devices \cite{Juanjuan_Optica_2019, Guo_Optica_2016}. Moreover, the maximal SHG power we obtained is $\approx$ 2.2~mW SHG in the waveguide with $\approx$ 15~mW input pump power (the bottom panel of Fig.~\ref{Fig4}(a)). Such milliwatt-level output SHG powers are a record among all nanophotonic SHG results~\cite{Levy_OE_2011, Zhang_NatPhoton_2019, Juanjuan_Optica_2019, Guo_Optica_2016}.

As discussed in the Supplementary Material, we estimate the induced effective second-order nonlinearity to be $\chi^{(2)}_\text{eff}$ = (0.20 $\pm$ 0.04)~pm/V. This value is near the lower end of previous Si$_3$N$_4$ photogalvanic results~\cite{Porcel_OE_2017, Billat_NatCommun_2017, Hickstein_NatPhoton_2019, Grassani_OL_2019, Edgars_ACSPhoton_2019}, which range from 0.3~pm/V to 3.7~pm/V. The electric field is estimated to be (0.6 $\pm$ 0.1)~MV/cm, which is $\approx$ 15~\% to 20~\% of the electric breakdown voltage of Si$_3$N$_4$~\cite{Rauthan_MaterLett_1990}, i.e., 3~MV/cm to 4~MV/cm. Our induced nonlinearity is currently limited by pump depletion and likely can be further improved. For example, in previous work~\cite{Timurdogan_NatPhoton_2017}, the applied field is 0.25~MV/cm, which is 62.5~\% of the electric breakdown voltage of Si (0.40~MV/cm).


In summary, we demonstrate efficient photo-induced second harmonic generation (SHG) with perfect-phase matching in silicon photonics, achieving record-high conversion efficiencies in comparison to prior silicon-based devices, and absolute efficiencies and output powers on par with the highest values demonstrated in nanophotonic media with much larger intrinsic $\chi^{(2)}$ nonlinearities. Our demonstration opens up promising avenues for $\chi^{(2)}$ nonlinear silicon photonics, including $f$-2$f$ locking for $\chi^{(3)}$-mediated octave-span frequency combs, sum-/difference-frequency generation, and electro-optical modulation.\\

\noindent \textbf{Acknowledgements} This work is supported by the DARPA ACES, DARPA DODOS, and NIST-on-a-chip programs. X.L. acknowledges support under the Cooperative Research Agreement between the University of Maryland and NIST-PML, Award no. 70NANB10H193.

\bibliographystyle{osajnl}

\begin{thebibliography}{10}
\newcommand{\enquote}[1]{``#1''}

\bibitem{Boyd2008}
R.~W. Boyd, \emph{{Nonlinear Optics}} (Academic Press, Amsterdam, 2008).

\bibitem{Spencer_Nature_2018}
D.~T. Spencer, T.~C. Drake, Tara~Briles, J.~Stone, L.~C. Sinclair, C.~Fredrick,
  Q.~Li, D.~Westly, B.~R. Ilic, A.~Bluestone, N.~Volet, T.~Komljenovic,
  L.~Chang, S.~H. Lee, D.~Y. Oh, M.-G. Suh, K.~Y. Yang, M.~H.~P. Pfeiffer,
  T.~J. Kippenberg, E.~Norberg, L.~Theogarajan, K.~Vahala, N.~R. Newbury,
  K.~Srinivasan, J.~E. Bowers, S.~A. Diddams, and S.~B. Papp, \enquote{{An
  optical-frequency synthesizer using integrated photonics},} Nature
  \textbf{557}, 81--85 (2018).

\bibitem{Singh_CLEO_2019}
N.~Singh, M.~Xin, D.~V. N.~Li, E.~S.~M. A.~Ruocco, K.~Shtyrkova, P.~T.
  Callahan, E.~Ippen, F.~X. K?rtner, , and M.~R. Watts, \enquote{{Silicon
  photonics optical frequency synthesizer - SPOFS},} Conference on Lasers and
  Electro-Optics p. ATh4I.2 (2019).

\bibitem{Newman_Optica_2019}
Z.~L. Newman, V.~Maurice, T.~Drake, J.~R. Stone, T.~C. Briles, D.~T. Spencer,
  C.~Fredrick, Q.~Li, D.~Westly, B.~R. Ilic, B.~Shen, M.-G. Suh, K.~Y. Yang,
  C.~Johnson, D.~M.~S. Johnson, L.~Hollberg, K.~J. Vahala, K.~Srinivasan, S.~A.
  Diddams, J.~Kitching, S.~B. Papp, and M.~T. Hummon, \enquote{Architecture for
  the photonic integration of an optical atomic clock,} Optica \textbf{6},
  680--685 (2019).

\bibitem{Guo_Optica_2016}
X.~Guo, C.-L. Zou, and H.~X. Tang, \enquote{{Second-harmonic generation in
  aluminum nitride microrings with 2500 \%/W conversion efficiency},} Optica
  \textbf{3}, 1126--1131 (2016).

\bibitem{Lin_APLPhoton_2019}
L.~Chang, A.~Boes, P.~Pintus, J.~D. Peters, M.~Kennedy, X.~Guo, N.~Volet,
  S.~Yu, S.~A. Diddams, S.~B. Papp, , and J.~E. Bowers, \enquote{{High
  efficiency SHG in heterogenous integrated GaAs ring resonators},} APL Photon.
  \textbf{4}, 036103 (201p).

\bibitem{Juanjuan_Optica_2019}
J.~Lu, J.~B. Surya, X.~Liu, A.~W. Bruch, Z.~Gong, Y.~Xu, and H.~X. Tang,
  \enquote{Periodically poled thin-film lithium niobate microring resonators
  with a second-harmonic generation efficiency of 250,000\%/w,} Optica
  \textbf{6}, 1455--1460 (2019).

\bibitem{Wang_Optica_2018}
C.~Wang, C.~Langrock, A.~Marandi, M.~Jankowski, M.~Zhang, B.~Desiatov, M.~M.
  Fejer, and M.~Lon?ar, \enquote{Ultrahigh-efficiency wavelength conversion in
  nanophotonic periodically poled lithium niobate waveguides,} Optica
  \textbf{5}, 1438--1441 (2018).

\bibitem{Luo_Optica_2018}
R.~Luo, Y.~He, H.~Liang, M.~Li, and Q.~Lin, \enquote{Highly tunable efficient
  second-harmonic generation in a lithium niobate nanophotonic waveguide,}
  Optica \textbf{4}, 1251--1258 (2018).

\bibitem{Lin_Optica_2016}
N.~V. L. W. J. P. J. E.~B. L.~Chang, Y.~Li, \enquote{Thin film wavelength
  converters for photonic integrated circuits,} Optica \textbf{3}, 531--535
  (2016).

\bibitem{Lin_OL_2017}
L.~Chang, M.~H.~P. Pfeiffer, N.~Volet, M.~Zervas, J.~D. Peters, C.~L.
  Manganelli, E.~J. Stanton, Y.~Li, T.~J. Kippenberg, and J.~E. Bowers,
  \enquote{Heterogeneous integration of lithium niobate and silicon nitride
  waveguides for wafer-scale photonic integrated circuits on silicon,} Opt.
  Lett. \textbf{42}, 803--806 (2017).

\bibitem{Noda_Optica_2019}
B.-S. Song, T.~Asano, S.~Jeon, H.~Kim, C.~Chen, D.~D. Kang, and S.~Noda,
  \enquote{Ultrahigh-q photonic crystal nanocavities based on 4h silicon
  carbide,} Optica \textbf{6}, 991--995 (2019).

\bibitem{Lukin_NatPhoton_2019}
D.~M. Lukin, C.~Dory, M.~A. Guidry, K.~Y. Yang, S.~D. Mishra, R.~Trivedi,
  M.~Radulaski, S.~Sun, D.~Vercruysse, G.~H. Ahn, and
  J.~Vu$\text{\u{c}}$kovi$\text{\'{c}}$,
  \enquote{{4H-silicon-carbide-on-insulator for integrated quantum and
  nonlinear photonics},} Nat. Photon. pp. doi:10.1038/s41566--019--0556--6
  (2019).

\bibitem{Levy_OE_2011}
J.~S. Levy, M.~A. Foster, A.~L. Gaeta, and M.~Lipson, \enquote{Harmonic
  generation in silicon nitride ring resonators,} Opt. Express \textbf{19},
  11415--11421 (2011).

\bibitem{Zhang_NatPhoton_2019}
X.~Zhang, Q.-T. Cao, Z.~Wang, Y.-x. Liu, C.-W. Qiu, L.~Yang, Q.~Gong, and Y.-F.
  Xiao, \enquote{Symmetry-breaking-induced nonlinear optics at a microcavity
  surface,} Nat. Photon. \textbf{13}, 21--24 (2019).

\bibitem{Timurdogan_NatPhoton_2017}
E.~Timurdogan, C.~V. Poulton, M.~J. Byrd, and M.~R. Watts, \enquote{Electric
  field-induced second-order nonlinear optical effects in silicon waveguides,}
  Nat. Photon. \textbf{11}, 200--206 (2017).

\bibitem{Porcel_OE_2017}
M.~A. Porcel, J.~Mak, C.~Taballione, V.~K. Schermerhorn, J.~P. Epping, P.~J.
  van~der Slot, and K.-J. Boller, \enquote{Photo-induced second-order
  nonlinearity in stoichiometric silicon nitride waveguides,} Opt. Express
  \textbf{25}, 33143--33159 (2017).

\bibitem{Billat_NatCommun_2017}
A.~Billat, D.~Grassani, M.~H.~P. Pfeiffer, S.~Kharitonov, T.~J. Kippenberg, and
  C.-S. Br{\`{e}}s, \enquote{Large second harmonic generation enhancement in
  $\text{Si}_\text{3}\text{N}_\text{4}$ waveguides by all-optically induced
  quasi-phase-matching,} Nat. Commun. \textbf{8}, 1016 (2017).

\bibitem{Hickstein_NatPhoton_2019}
D.~D. Hickstein, D.~R. Carlson, H.~Mundoor, J.~B. Khurgin, K.~Srinivasan,
  D.~Westly, A.~Kowligy, I.~I. Smalyukh, S.~A. Diddams, and S.~B. Papp,
  \enquote{Self-organized nonlinear gratings for ultrafast nanophotonics,} Nat.
  Photon. \textbf{13}, 494--499 (2019).

\bibitem{Grassani_OL_2019}
D.~Grassani, M.~H.~P. Pfeiffer, T.~J. Kippenberg, and C.-S. Br¨¨s,
  \enquote{Second- and third-order nonlinear wavelength conversion in an
  all-optically poled {Si$_3$N$_4$} waveguide,} Opt. Lett. \textbf{44},
  106--109 (2019).

\bibitem{Edgars_ACSPhoton_2019}
E.~Nitiss, T.~Liu, D.~Grassani, M.~Pfeiffer, T.~J. Kippenberg, and C.-S.
  Br$\text{\`{e}}$s, \enquote{Formation rules and dynamics of photoinduced
  $\chi^{(2)}$ gratings in silicon nitride waveguides,} ACS Photon. p.
  10.1021/acsphotonics.9b01301 (2019).

\bibitem{Moss_NatPhoton_2013}
D.~J. Moss, R.~Morandotti, A.~L. Gaeta, and M.~Lipson, \enquote{New
  cmos-compatible platforms based on silicon nitride and hydex for nonlinear
  optics,} Nat. Photon. \textbf{7}, 597--607 (2013).

\bibitem{Okawachi_OL_2011}
Y.~Okawachi, K.~Saha, J.~S. Levy, Y.~H. Wen, M.~Lipson, and A.~L. Gaeta,
  \enquote{{Octave-spanning frequency comb generation in a silicon nitride
  chip.}} Opt. Lett. \textbf{36}, 3398--3400 (2011).

\bibitem{Li_Optica_2017}
Q.~Li, T.~C. Briles, D.~A. Westly, T.~E. Drake, J.~R. Stone, B.~R. Ilic, S.~A.
  Diddams, S.~B. Papp, and K.~Srinivasan, \enquote{{Stably accessing
  octave-spanning microresonator frequency combs in the soliton regime},}
  Optica \textbf{4}, 193--203 (2017).

\bibitem{Karpov_NatCommun_2018}
M.~Karpov, M.~H. Pfeiffer, J.~Liu, A.~Lukashchuk, and T.~J. Kippenberg,
  \enquote{{Photonic chip-based soliton frequency combs covering the biological
  imaging window},} Nat. Commun. \textbf{9}, 1146 (2018).

\bibitem{Li_NatPhoton_2016}
Q.~Li, M.~Davan{\c{c}}o, and K.~Srinivasan, \enquote{{Efficient and low-noise
  single-photon-level frequency conversion interfaces using silicon
  nanophotonics},} Nat. Photon. \textbf{10}, 406--414 (2016).

\bibitem{Lu_NatPhoton_2019}
X.~Lu, G.~Moille, Q.~Li, D.~A. Westly, A.~Rao, S.-P. Yu, T.~C. Briles, S.~B.
  Papp, and K.~Srinivasan, \enquote{{Efficient telecom-to-visible spectral
  translation using silicon nanophotonics},} Nat. Photon. \textbf{13}, 593--601
  (2019).

\bibitem{Lu_Optica_2019}
X.~Lu, G.~Moille, A.~Singh, Q.~Li, D.~A. Westly, A.~Rao, S.-P. Yu, T.~C.
  Briles, S.~B. Papp, and K.~Srinivasan, \enquote{{Milliwatt-threshold
  visible-telecom opticalparametric oscillation using silicon nanophotonics},}
  Optica \textbf{6}, 1535--1541 (2019).

\bibitem{UO_OL_1986}
U.~\"{O}sterberg and W.~Margulis, \enquote{Dye laser pumped by {Nd}:{YAG} laser
  pulses frequency doubled in a glass optical fiber,} Opt. Lett. \textbf{11},
  516--518 (1986).

\bibitem{UO_OL_1987}
U.~\"{O}sterberg and W.~Margulis, \enquote{Experimental studies on efficient
  frequency doubling in glass optical fibers,} Opt. Lett. \textbf{12}, 57--59
  (1987).

\bibitem{Tom_OL_1987}
R.~H. Stolen and H.~W.~K. Tom, \enquote{Self-organized phase-matched harmonic
  generation in optical fibers,} Opt. Lett. \textbf{12}, 585--587 (1987).

\bibitem{Tom_OL_1988}
H.~W.~K. Tom, R.~H. Stolen, G.~D. Aumiller, and W.~Pleibel,
  \enquote{Preparation of long-coherence-length second-harmonic-generating
  optical fibers by using mode-locked pulses,} Opt. Lett. \textbf{13}, 512--514
  (1988).

\bibitem{Margulis_Nature_1995}
W.~Margulis, F.~Laurell, and B.~Lesche, \enquote{Imaging the nonlinear grating
  in frequency-doubling fibres,} Nature \textbf{378}, 699--701 (1995).

\bibitem{Ning_OL_2012}
T.~Ning, H.~Pietarinen, O.~Hyv?rinen, R.~Kumar, T.~Kaplas, M.~Kauranen, and
  G.~Genty, \enquote{Efficient second-harmonic generation in silicon nitride
  resonant waveguide gratings,} Opt. Lett. \textbf{37}, 4269--4271 (2012).

\bibitem{Lu_APL_2014}
X.~Lu, S.~Rogers, W.~C. Jiang, and Q.~Lin, \enquote{{Selective engineering of
  cavity resonance for frequency matching in optical parametric processes},}
  Appl. Phys. Lett. \textbf{105}, 151104 (2014).

\bibitem{Rauthan_MaterLett_1990}
C.~Rauthan and J.~Srivastava, \enquote{Electrical breakdown voltage
  characteristics of buried silicon nitride layers and their correlation to
  defects in correlation to defects in the nitride layer,} Mater. Lett.
  \textbf{9}, 252--258 (1990).

\bibitem{Ikeda_OE_2008}
K.~Ikeda, R.~E. Saperstein, N.~Alic, and Y.~Fainman, \enquote{{Thermal and Kerr
  nonlinear properties of plasma-deposited silicon nitride/silicon dioxide
  waveguides},} Opt. Express \textbf{16}, 12987--12994 (2008).

\end{thebibliography}

\newpage
\begin{thebibliography}{10}
\newcommand{\enquote}[1]{``#1''}

\bibitem{coimbatore_balram_nanolithography_2016}
K.~C. Balram, D.~A. Westly, M.~I. Davanco, K.~E. Grutter, Q.~Li, T.~Michels,
  C.~H. Ray, R.~J. Kasica, C.~B. Wallin, I.~J. Gilbert, B.~A. Bryce,
  G.~Simelgor, J.~Topolancik, N.~Lobontiu, Y.~Liu, P.~Neuzil, V.~Svatos, K.~A.
  Dill, N.~A. Bertrand, M.~Metzler, G.~Lopez, D.~Czaplewski, L.~Ocola, K.~A.
  Srinivasan, S.~M. Stavis, V.~A. Aksyuk, J.~A. Liddle, S.~Krylov, and B.~R.
  Ilic, \enquote{The nanolithography toolbox,} Journal of Research of the
  National Institute of Standards and Technology \textbf{121}, 464--475 (2016).

\end{thebibliography}

\newpage

\begin{widetext}

\section*{\large{\textbf{Supplementary Information}}}

\section{A toy model for comparison purposes}
In this section, we review a simple model to understand the differences in efficiency of various approaches to SHG, as illustrated in Fig.~\ref{FigS1}(a) (same as Fig.~1(c) in the main text). This toy model compares the SHG response from cavities and waveguides in perfect-/quasi-phase matching (PPM/QPM) cases.

As shown in Fig.~\ref{FigS1}(b), the effective nonlinearity in our case is always maximal, as it is induced by a DC field with perfect-phase matching (solid red). For photo-induced SHG with perfect phase-matching in a waveguide, because the build-up of the photogalvanic effect requires certain thresholds for the input light fields, the nonlinearity needs some time to be turned on (solid blue line in Fig~\ref{FigS1}(b)). In the quasi-phase matching cases, for both cavity (dashed red line) and waveguide (dashed blue line), the effective nonlinearity has a sinusoidal modulation compared to the perfect-phase matching case. In the waveguide (dashed blue line), this modulation also leads to a slower response in the build-up of the nonlinearity. Finally, any seedling nonlinearity ($\chi^{(2)}_\text{s}$) is much smaller than the effective/induced nonlinearity ($\chi^{(2)}_\text{eff}$); we assume $\chi^{(2)}_\text{s} = 0.1~\chi^{(2)}_\text{eff}$ here (solid purple line). Here the decay-down of quasi-phase matching in the propagation direction, a practical issue for waveguide/fiber systems, is not considered. Moreover, cavity enhancement in time is not considered in this simple model (the cavity Purcell effect provides a factor of $\approx~5,000$ speed-up in time).

\begin{center}
\begin{figure*}[h!]
\begin{center}
\includegraphics[width=0.9\linewidth]{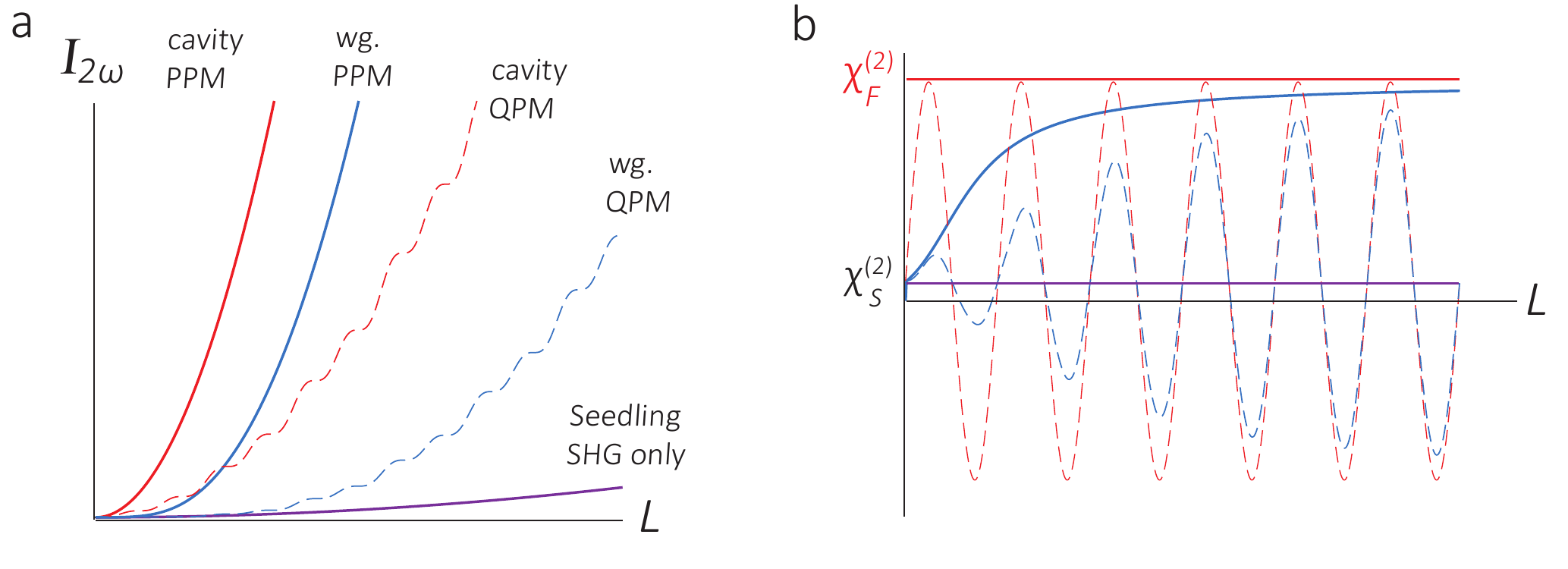}
  \caption{\textbf{Toy model for comparison.} \textbf{a,} The SHG response in the toy model for a cavity and waveguide in both perfect-/quasi-phase matching (PPM/QPM) cases. The cavity PPM case (solid red) is the best among all the schemes. \textbf{b,} A comparison of the build-up in the nonlinearities for the corresponding schemes, with the same color scheme as shown in (a). The cavity PPM case is most effective because the field-induced nonlinearity is the largest and remains constant in the propagation direction.}
\label{FigS1}
\end{center}
\end{figure*}
\end{center}
The equations that describe the build-up of the nonlinearity in these systems within this simple model are given as follows:
\begin{eqnarray*}
\chi^{(2)} =
    \begin{cases}
      ~~~\chi_F^{(2)} & \text{cavity PPM},\\
      ~~~\chi_F^{(2)}/(1+E_\text{th}/E_{2\omega}) +\chi_S^{(2)} & \text{wg. PPM},\\
      ~~~\chi_F^{(2)}~sin(2 \pi L/a) & \text{cavity QPM},\\
      ~~~\chi_F^{(2)}~sin(2 \pi L/a)/ (1+E_\text{th}/E_{2\omega}) +\chi_S^{(2)} & \text{wg. QPM},\\
      ~~~\chi_S^{(2)} & \text{seedling SHG only},\\
    \end{cases}
    \label{EqS1}
\end{eqnarray*}
where $\chi_F^{(2)}$ is the field-induced second-order nonlinearity, i.e., $\chi_F^{(2)} = \chi^{(3)} E_{DC}$. $\chi_S^{(2)}$ is the small seedling SHG nonlinearity. For illustration purposes, we use $\chi_S^{(2)} = 0.1~\chi_F^{(2)}$ here. $E_\text{th}$ represents the threshold electric field amplitude to excite the photo-galvanic effect and $a$ represents the period of quasi-phase matching.

\section{A full model for photo-galvanic DC-field-induced SHG}
\begin{center}
\begin{figure}[h!]
\begin{center}
\includegraphics[width=0.9\linewidth]{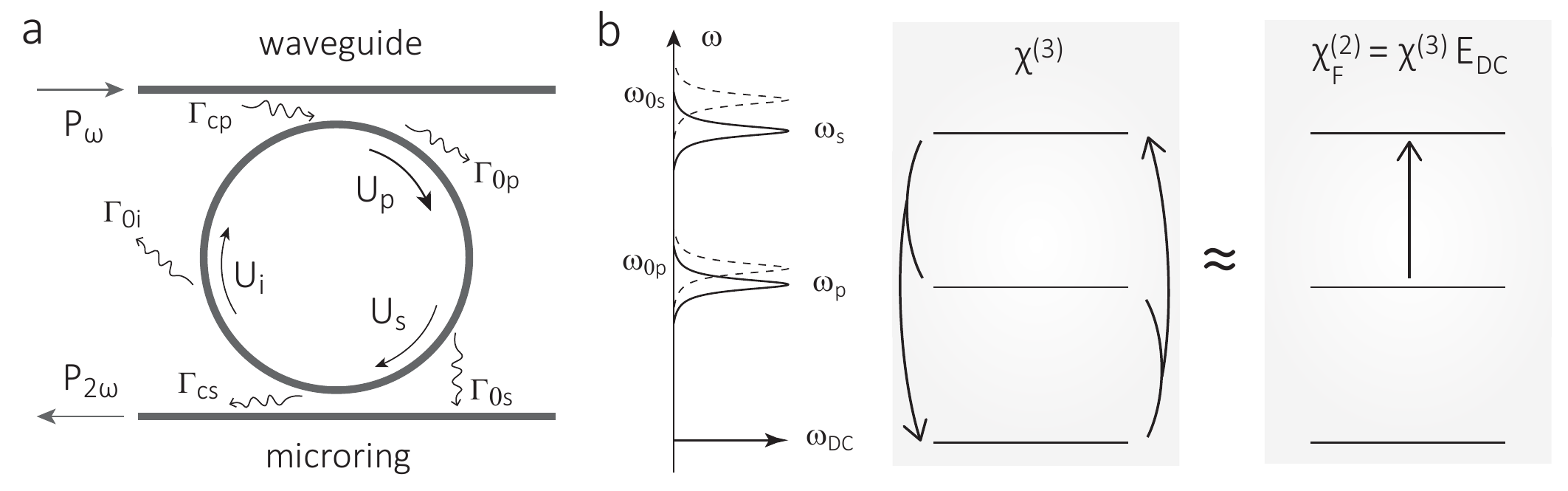}
  \caption{\textbf{Illustration of device parameters and effective second-order nonlinearity.} \textbf{a,} Device blueprint for SHG. Three fields, the pump field, SHG field, and DC field, are present inside the microring. $U_\text{m}$ represents intra-cavity energy, where $m$ = $p$, $s$, $i$ represent pump, SHG, DC fields, respectively. The intrinsic loss (from absorption and scattering) of each field is described by $\Gamma_\text{0m}$. Two waveguides are used to couple pump (top) and SHG light (bottom) with rates of $\Gamma_\text{cp}$ and $\Gamma_\text{cs}$, respectively. \textbf{b,} When the pump and SHG modes are matched in frequency, the DC Kerr effect, a third-order nonlinear process, is equivalent to SHG, a second-order nonlinear process. The effective second-order nonlinearity is the product of the third-order nonlinearity and the induced DC field.}
\label{FigS2}
\end{center}
\end{figure}
\end{center}

In this section, we review second harmonic generation (SHG) in high-$Q$ microresonators, and present an estimate of related physical parameters, including mode overlap ($\bar{\eta}$), effective mode volumes ($\bar{V}$), induced nonlinearity ($\chi^{(2)}$), and electric field ($E_\text{DC}$). In particular, we look into the cases where pump and SHG modes are quite different in frequency. In high-Q microresonators, because light propagates many round trips before being lost (e.g., scattering or absorption) or appreciably coupled out from the cavity, we can treat the loss and coupling as if they are uniformly distributed in time and space. The slowly varying light fields satisfy the following equations given by:
\begin{eqnarray}
\frac{d\tilde{A}_\text{p}}{dt} &=& (i \Delta\omega_\text{p} - \Gamma_\text{tp}/2) \tilde{A}_\text{p} + i \gamma \tilde{A}_\text{s} \tilde{A}^*_\text{p} + i \Gamma^{1/2}_\text{cp} \tilde{S}_\text{p}, \label{EqS2} \\
\frac{d\tilde{A}_\text{s}}{dt} &=& (i \Delta\omega_\text{s} - \Gamma_\text{ts}/2) \tilde{A}_\text{s} + 2 i \gamma \tilde{A}^2_\text{p}, \label{EqS3}
\end{eqnarray}
where $\tilde{A}_\text{m}$ (m = p,s) are the intra-cavity light fields for pump and SHG modes, sitting on the fast-oscillating background of $e^{-i\omega_m t}$, where $\omega_m$ is the angular frequency of the light. Frequency conservation requires $\omega_s=2\omega_p$, which is assumed in deducing the equations. The cavity fields are normalized so that  $|\tilde{A}_\text{m}|^2$ = $U_\text{m}$ (m = p,s), which represents the intra-cavity energy. The first terms in Eqs.~(\ref{EqS2}-\ref{EqS3}) describe the free cavity evolution (without sources or nonlinear effects), where $\Delta\omega_\text{m}$ (m = p,s) represents the detuning of laser/light frequency ($\omega_\text{m}$) from the natural cavity frequency ($\omega_\text{0m}$), i.e., $\Delta\omega_\text{m}=\omega_\text{m}-\omega_\text{0m}$. $\Gamma_\text{tm}$ describes the decay of the intra-cavity energy $U_\text{m}$, which includes the intrinsic cavity loss and the out-coupling to waveguide, $\Gamma_\text{tm} = \Gamma_\text{0m} +\Gamma_\text{cm}$. Here the decay term $\Gamma_\text{lm}$ is related to optical quality factor $Q_\text{lm}$ by:
\begin{eqnarray}
\Gamma_\text{lm} = \frac{\omega_\text{0m}}{Q_\text{lm}},~(l = t,0,c;~m = p,s). \label{EqS4}
\end{eqnarray}
We use $\Gamma$ instead of $Q$ so that it is more straightforward to describe the physics of the cavity, as shown in Fig.~\ref{FigS2}(a).

\begin{center}
\begin{figure}[h!]
\begin{center}
\includegraphics[width=0.9\linewidth]{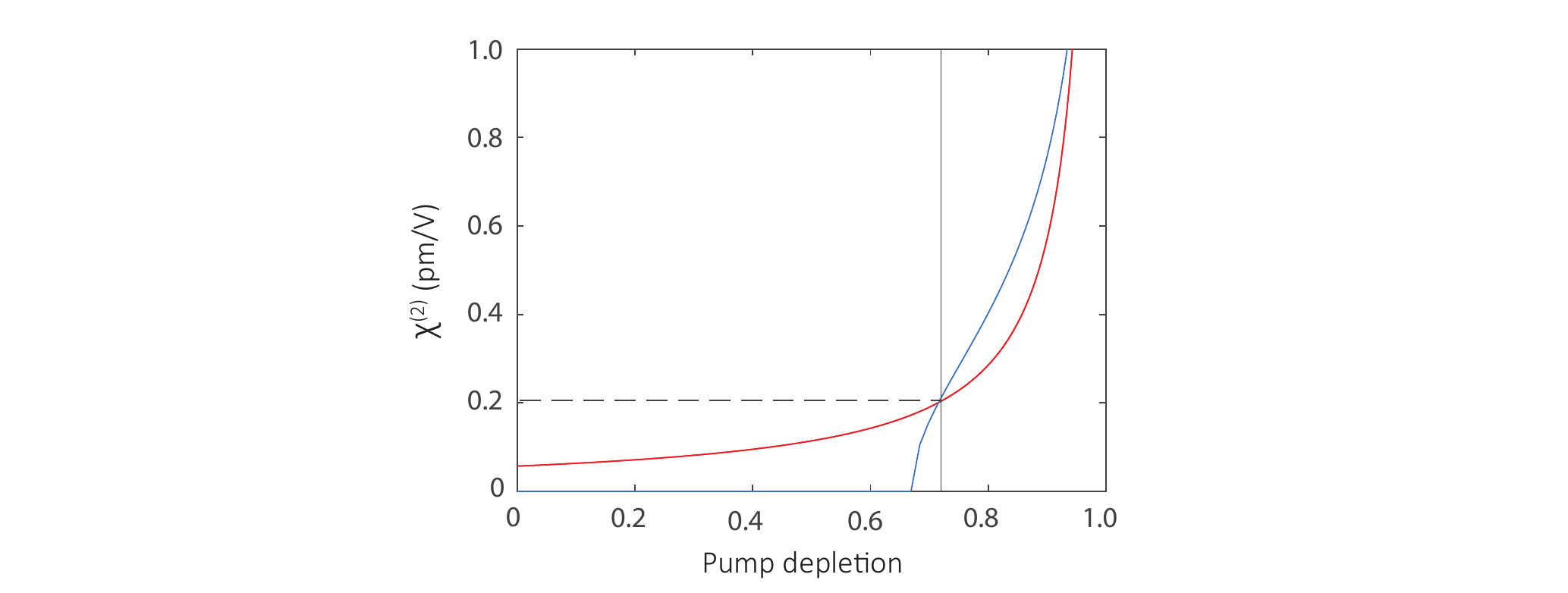}
  \caption{\textbf{Estimate of the photo-induced $\chi^{(2)}$.} The inferred nonlinearity as a function of pump depletion is given by the red line (Eq.~\ref{EqS9}). In our work, the pump is deeply depleted by $\approx$ 72 \% (Section V), indicated by the vertical black line. The intersection of the black and red lines gives a solution of (0.20 $\pm$ 0.04) pm/V (dashed black line) for the second-order nonlinearity. The error is propagated from the one-stand derivation uncertainties in conversion efficiencies and optical quality factors. The blue line represents a plot of Eq.~\ref{EqS8} with $\Delta\omega_\text{p}$ $\approx$ $\Gamma_\text{tp}$.}
\label{FigS3}
\end{center}
\end{figure}
\end{center}
The second terms in Eqs.~(\ref{EqS2}-\ref{EqS3}) describe the SHG processes. Here the nonlinear parameter $\gamma$ is given by
\begin{eqnarray}
\gamma = \frac{3 \omega_\text{p} \bar{\eta} \chi_\text{F}^{(2)}}{4 \sqrt{2} \bar{n}^3 \sqrt{\epsilon_\text{0} \bar{V}}}, \label{EqS5}
\end{eqnarray}
where $\bar{\eta}$ characterizes the spatial overlap of interacting optical modes and is therefore a dimensionless real parameter from 0 to 1, with its value given by:
\begin{eqnarray}
\bar{\eta} = \frac{\int_\text{V}dv~\sqrt{\epsilon_\text{s}} \epsilon_\text{p} \tilde{E}^*_\text{s} \tilde{E}^2_\text{p}} {(\int_\text{V}dv~\epsilon^{3/2}_\text{p} {|\tilde{E}_\text{p}|}^3)^{2/3} (\int_\text{V}dv~\epsilon^{3/2}_\text{s} {|\tilde{E}_\text{s}|}^3)^{1/3}}, \label{EqS6}
\end{eqnarray}
Here $\tilde{E}_\text{m}$ represents the dominant electric field components of the m = p,s mode. This field is related to $A_\text{m}$ by $U_\text{m} = |A_\text{m}|^2 \approx \int_\text{V}dv~\epsilon_\text{m} |\tilde{E}_\text{m}|^2$. Here the approximation is made possible when the other electric field components are much smaller than the dominant one, for example, $|\tilde{E}_\text{z}|$, $|\tilde{E}_{\phi}|$ $\ll$ $|\tilde{E}_\text{r}|$ for transverse-electric (TE) modes. The mode overlap is estimated to be 67.6 \% using finite-element method simulation. In Eq.~(\ref{EqS5}), $\chi_\text{F}^{(2)} = \chi^{(3)} E_\text{DC}$, where $\chi^{(3)}$ is short for $\chi^{(3)}(-\omega_\text{s}; \omega_\text{p}, -\omega_\text{p}, 0)$ and represents the third-order nonlinearity at $\omega_\text{s}$ with the inputs at $\omega_\text{p}$, $\omega_\text{p}$, and a DC field. $\tilde{E}_\text{DC}$ is the equivalent DC field amplitude from the photo-galvanic effect, assuming that such a field is evenly distributed inside the microring and that the mode volume and the mode overlap of $\chi^{(2)}$ and $\chi^{(3)}$ are similar. $\bar{n}$ represents average linear refractive index $\bar{n} = (n_\text{p}^2 n_\text{s})^{1/3}$. Likewise, $\bar{V}$ represents average mode volume $\bar{V} = (V^2_\text{p} V_\text{s})^{1/3}$, where individual mode volume is given by:
\begin{eqnarray}
V_\text{m} = \frac{{({\int_\text{V} dv~\epsilon_\text{m}|\tilde{E}_\text{m}|}^2)}^{3}}{{(\int_\text{V}dv~\epsilon^{3/2}_\text{m} {|\tilde{E}_\text{m}|}^3)}^2},~\text{(with m = p,s).} \label{EqS7}
\end{eqnarray}
We use finite-element method simulation to calculate the mode volumes of the pump and SHG modes to be 60.6 $\mu$m$^3$ and 54.0 $\mu$m$^3$. The effective mode volume is therefore 58.3 $\mu$m$^3$.
Note that the nonlinear parameter $\gamma$ described in Eq.~\ref{EqS5} is related to nonlinear coupling strength \cite{Guo_Optica_2016}, i.e., $g$, with a normalization of photon energy given by $\gamma = g/ \sqrt{\hbar \omega_\text{s}}$.

The last term in Eq.~(\ref{EqS2}) is the source term that represents the pump laser coupled into the cavity. The coupling rate $\Gamma_\text{cp}$ is given by Eq.~\ref{EqS4} and the input field $\tilde{S}_\text{in}$ is normalized in such a way that $|\tilde{S}_\text{p}|^2 = P_{\omega}$ represents the input power of the pump laser in the waveguide (Fig.~\ref{FigS2}).

We note that terms representing phenomena such as nonlinear absorption and free carrier effects are not considered in Eqs.~(\ref{EqS2}-\ref{EqS3}), as silicon nitride (Si$_3$N$_4$) is wide-bandgap and does not have such effects in the frequency ranges of interest in this work. In addition, four-wave mixing (FWM) of the cavity fields is not considered in these equations. Self/cross-phase modulation and the thermo-optical shift, although not considered explicitly, can be included in the detuning of the pump and SHG modes ($\Delta\omega_\text{p}$ and $\Delta\omega_\text{s}$). See Section IV for details. Quantum fluctuation of the pump and signal bands is also not included because we are only interested in the classical regime. When considered in steady-state in the continuous-wave case, Eqs.~(\ref{EqS2})-(\ref{EqS3}) reduce to:
\begin{eqnarray}
~[\Delta\omega^2_\text{p}+(\Gamma_\text{tp}/2)^2] U_\text{p}+\gamma^2 U_\text{p} U_\text{s} = \Gamma_\text{cp} P_{\omega}, \label{EqS8} \\
~[\Delta\omega^2_\text{s}+(\Gamma_\text{ts}/2)^2] U_\text{s} = 4 \gamma^2 U^2_\text{p}, \label{EqS9}
\end{eqnarray}

In the perturbative regime, the second term in Eq.~(\ref{EqS8}) can be neglected and system response is linearized, therefore resulting in the SHG efficiency:
\begin{eqnarray}
\eta \equiv P_{2\omega}/P^2_{\omega} = 4 \gamma^2 \frac{\Gamma_\text{cs}}{\Delta\omega^2_\text{s}+(\Gamma_\text{ts}/2)^2} \frac{\Gamma^2_\text{cp}}{{[\Delta\omega^2_\text{p}+(\Gamma_\text{tp}/2)^2]}^2}, \label{EqS10}
\end{eqnarray}

When the pump is depleted, that is, $\gamma \gtrapprox \Gamma_\text{tp}/A_\text{s}$, Eqs.~(\ref{EqS8})-(\ref{EqS9}) lead to a nonlinear dependence on $U_\text{p}$. We solve this equation graphically in Fig.~\ref{FigS3}, and retrieve a $\chi_\text{F} = (0.20~\pm~0.04)$ pm/V, where the error is propagated from the uncertainties in conversion efficiencies and optical quality factors. This value corresponds to a DC field of $(0.6~\pm~0.1)$ MV/cm, using $\chi^{(3)} = 3.39 \times 10^{-21} \text{m}^2/\text{V}^2$ as the third-order nonlinearity~\cite{Ikeda_OE_2008}. The error is propagated from the one-standard-deviation uncertainty in estimating $\chi_\text{F}$. This estimation assumes that pump and SHG fiber-chip coupling and laser detuning (i.e., frequency matching) are all optimized in the experiment.

\bigskip \section{Device simulation and fabrication}
In this section, we provide extra data for the device dispersion and coupling. In the main text, the two optical modes used in the experiment have radial and azimuthal mode numbers of (1, 154) for the pump mode and (3, 308) for the SHG mode. We calculate the frequency mismatch of these two modes by finite-element method simulation, as shown in Fig.~\ref{FigS4}(b), where the ring width ($RW$) and thickness ($H$) vary while the ring radius ($RR$) is fixed. The nominal device structure has $RW$ = 1200 nm, $H$ = 600 nm, and $RR$ = 23 $\mu$m and its frequency mismatch $\Delta \nu = \nu_\text{s} -2 \nu_\text{p}$ = -150~GHz. This frequency mismatch is sensitive to device geometries (e.g., $RW$ and $H$) and can be tuned by rates of $\delta (\Delta\nu)/\delta RW \approx $  44 GHz/nm and $\delta (\Delta \nu)/\delta H \approx $  50 GHz/nm. The device used in the experiments has a frequency mismatch of 7.7~GHz, which is small enough to be compensated for in practice via thermal and Kerr effects (Section IV). Two waveguides are used to couple the pump and SHG modes separately, as shown in Fig.~\ref{FigS4}(a). The nominal device parameters of the coupling waveguides are given in (c). The nominal parameters yield coupling $Q$~$\approx$~10$^6$ for both pump (left) and SHG (right) modes, as shown in Fig.~\ref{FigS4}(d).
\begin{center}
\begin{figure}[h!]
\begin{center}
\includegraphics[width=0.9\linewidth]{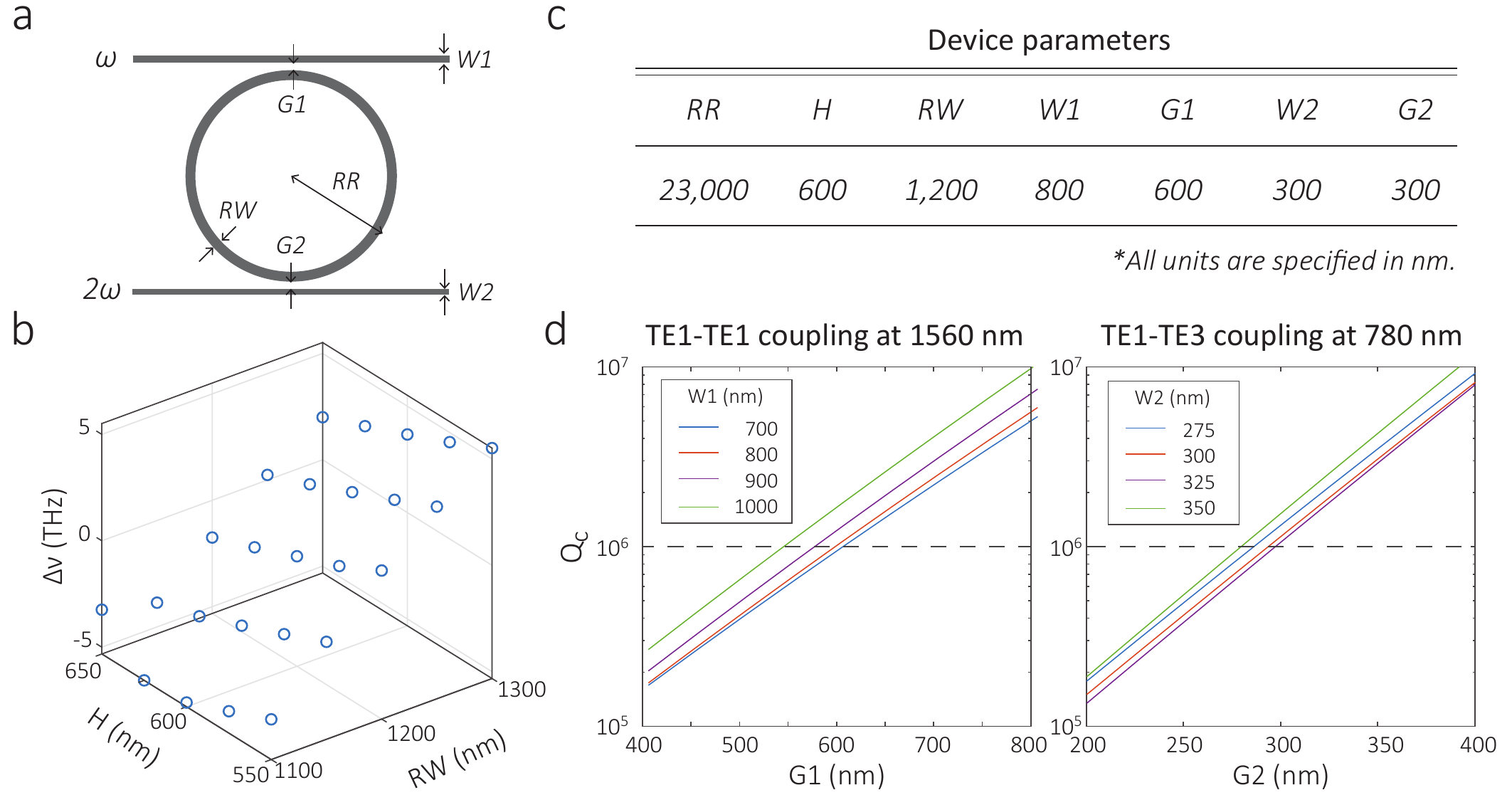}
  \caption{\textbf{Device parameters.} \textbf{a,} Schematic of the SHG device. Two waveguides are used to couple pump and SHG light separately. Three parameters control the dispersion of the microring: thickness ($H$), microring width ($RW$), and microring radius ($RR$). Two parameters are needed to define the coupling to the straight waveguide (wg.): waveguide width ($W$) and gap ($G$). \textbf{b,} The simulated frequency mismatch ($\Delta \nu$) of pump and SHG modes with mode numbers of (1, 154) and (3, 308), respectively. The device with $H$ = 600 nm and $RW$ = 1200 nm has the smallest frequency mismatch of -150~GHz. \textbf{c,} A parameter table for the typical geometries studied in the main text. \textbf{d,} The simulated gap-dependent coupling with various waveguide widths for both pump (1560 nm) and SHG (780 nm) light.}
\label{FigS4}
\end{center}
\end{figure}
\end{center}

The device layout was done with the Nanolithography Toolbox, a free software package developed by the NIST Center for Nanoscale Science and Technology~\cite{coimbatore_balram_nanolithography_2016}. The ${\rm Si_3N_4}$ layer is deposited by low-pressure chemical vapor deposition on top of a 3~${\rm \mu}$m thick thermal ${\rm SiO_2}$ layer on a 100~mm diameter Si wafer. The wavelength-dependent refractive index and the thickness of the layers are measured using a spectroscopic ellipsometer, with the data fit to an extended Sellmeier model. The device pattern is created in positive-tone resist by electron-beam lithography. The pattern is then transferred to ${\rm Si_3N_4}$ by reactive ion etching using a ${\rm CF_4/CHF_3}$ chemistry. The device is chemically cleaned to remove deposited polymer and remnant resist, and then annealed at 1100~${\rm ^{\circ} C}$ in a ${\rm N_2}$ environment for 4 hours. An oxide lift-off process is performed so that the microrings have an air cladding on top while the input/output edge-coupler waveguides have ${\rm SiO_2}$ on top to form more symmetric modes for coupling to optical fibers. The facets of the chip are then polished for lensed-fiber coupling. After being polished, the chip is annealed again at 1100~${\rm ^{\circ} C}$ in a ${\rm N_2}$ environment for 4 hours.

\bigskip \section{Thermal shift}

We measure the thermal shift of the pump and SHG modes by temperature tuning as shown in Fig.~\ref{FigS5}. At room temperature ($\approx$ 21.9 $^\text{o}$C), the SHG wavelength (778.877~nm) is 15.5~pm smaller than half of the pump wavelength (1557.785~nm). However, the thermal shift rate of the SHG mode (11.11~pm/$^\text{o}$C) is larger than half of the rate of the pump mode (10.525~pm/$^\text{o}$C). Therefore, this mismatch can be compensated  at a rate of 0.585~pm/$^\text{o}$C by heating the device. Thus, at 46.4~$^\text{o}$C, these two modes are matched in the cold cavity. Here we only consider the frequency shift through temperature tuning in the cold cavity case, that is, the optical power causes no thermo-optic bistability. In the pumped cavity, the optical power creates heat inside the microring by increasing the temperature of the device locally. Moreover, the optical power also induces a Kerr shift, which is larger for cross-phase modulation on the SHG mode than self-phase modulation on the pump mode, and also helps compensate the frequency mismatch. In the experiment, as shown in Fig.~3(a), the cavity is found to match well at $\approx$ 1558.10 nm, with temperature of 27.8~$^\text{o}$C, where both thermal and Kerr effects contribute to realizing frequency matching.
\begin{center}
\begin{figure}[h!]
\begin{center}
\includegraphics[width=0.9\linewidth]{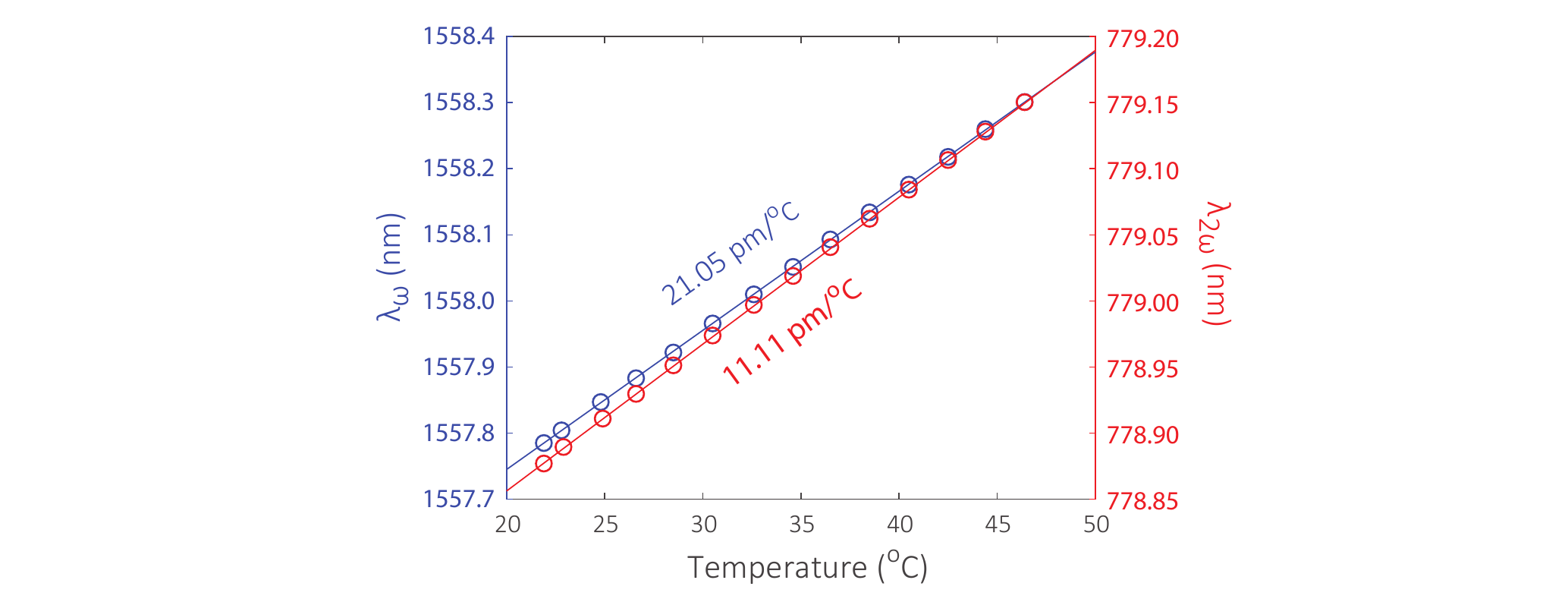}
  \caption{\textbf{Thermal shift to compensate frequency mismatch.} At room temperature ($\approx$ 21.9 $^\text{o}$C), the SHG wavelength is $\approx$ 15.5~pm smaller than half of the pump wavelength. The thermal shifts of the SHG mode and pump mode are different, so that at 46.4~$^\text{o}$C, these two modes can match (within 1 pm). The measurements are done in the cold cavity case, where the optical power is small and causes no thermo-optic or Kerr shifts.}
\label{FigS5}
\end{center}
\end{figure}
\end{center}

\section{Additional data}
\begin{center}
\begin{figure}[h!]
\begin{center}
\includegraphics[width=0.9\linewidth]{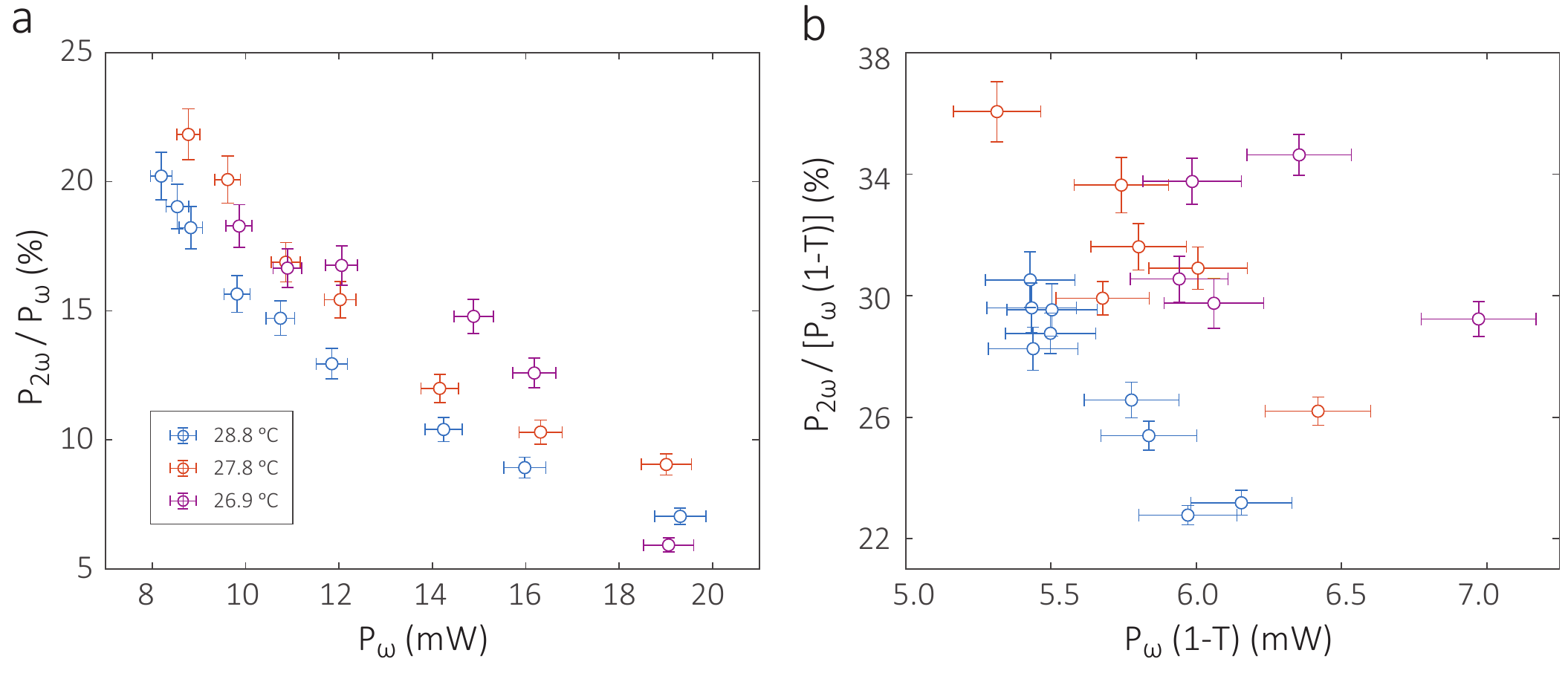}
  \caption{\textbf{Absolute SHG conversion efficiencies.} \textbf{a,} Absolute SHG conversion efficiency, determined by the ratio of SHG power versus pump power in the waveguides on chip. \textbf{b,} Absolute SHG efficiency normalized to dropped pump power into the microring. The errobars in (a) represent one-standard deviation uncertainties from the calibration of the on-chip power.}
\label{FigS6}
\end{center}
\end{figure}
\end{center}
In the main text, we show the comparison of SHG conversion efficiency normalized by the pump power. We show here the absolute conversion efficiency in the experiment, as well as the conversion efficiency normalized to dropped pump power (i.e., pump power coupled into the cavity). Each data point is optimized in pump detuning by the backwards tuning method described in the main text. The absolute conversion efficiency, that is, SHG power versus pump power in the waveguide (on chip), reaches up to $\approx$ 22~\%, as shown in Fig.~\ref{FigS6}(a). As shown in Fig.~\ref{FigS6}(b), the conversion efficiency normalized to the dropped pump power is as high as $\approx$ 36 \%. Thus up to $\approx$ 72 \% of the pump light is depleted inside the cavity, as the SHG mode is close to critical coupling. Such depletion explains why the optimization requires backward tuning, and also why it is not stable beyond this optimized point when tuned further in the backward direction.

\end{widetext}

\end{document}